\documentstyle[11pt,aas2pp4,flushrt,epsf]{article}    

\begin{document}

\title{A Note on the Statistical Mechanics of 
Violent Relaxation of Phase Space Elements of Different Densities}
\author{A. Kull, R. A. Treumann and H. B\"ohringer}
\affil{Max-Planck-Institut f\"ur Extraterrestrische Physik\\
        D-85740 Garching, Germany}

\begin{abstract}
The statistical mechanical investigation of Violent Relaxation of phase space
elements of different densities first derived by Lynden-Bell (1967) is
re-examined. It is found that the mass independence of the equations  of motion
of Violent Relaxation calls for a constraint on the volume of  the phase space
elements used to formulate the statistical mechanical description  of Violent
Relaxation. In agreement with observations of astrophysical  objects believed
to have been subject to Violent Relaxation (e.g. clusters  of galaxies), the
coarse grained phase space distribution $\bar{f}$  of the final state in the
non-degenerate limit turns into a  superposition of Maxwellians
of a common velocity dispersion. Thus, the velocity dispersion problem present
in the investigation of  Lynden-Bell (1967) is removed.
\end{abstract}


\section{INTRODUCTION}
Since the classical paper by Lynden-Bell (1967), the  statistical investigation
of Violent Relaxation has been  a topic of crucial interest in the discussion
of the formation of relaxed astrophysical objects like star clusters, galaxies
and clusters of galaxies. Its  reformulation by Shu (1978) in terms of
particles instead of  phase space elements represents a change in terminology
and, to some extent, a re-interpretation of the Liouville theorem.  However,
the basic result of both attempts is essentially the same and has been a
subject of continuous discussion.  The debate of Madsen (1987) and Shu (1987)
concerning the  foundations of a statistical description of Violent Relaxation 
provides a summary of resolved and unresolved issues for both the particle and
the phase-space representation.

Starting with Tremaine, Henon, \& Lynden-Bell (1986), the form and  evolution
of the $H$-function during Violent Relaxation became a  subject of growing
interest. Ever since one of the main lines  along which the statistical
mechanical investigation of violent relaxation proceeds, deals with this
problem (e.g., Stiavelli  \& Bertin 1987; Kandrup 1987). Only  recently (Soker
1996) it became clear, that the $H$-function of collisionless self-gravitating
systems does not necessarily increase monotonically during the time evolution
of the system. 

In the present paper we deal with another issue central for both  the particle
and the phase-space description of Violent Relaxation. Since the very first
investigation by Lynden-Bell (1967), it is considered as one of the most
important flaws of the theory  (Madsen 1987; Shu 1987) that despite the
collisionless nature of  Violent Relaxation the statistical mechanical approach
predicts a thermalized final state, i.e. a state in which the square of the
velocity dispersion  of phase space elements of different densities is
inversely proportional to the  density. In order to cure this problem, Shu
(1987) applied  very stringent assumptions on the initial phase space density
to  his particle approach. 

Here we re-examine the statistical mechanics of Violent Relaxation in terms  of
phase space elements of different densities. Since the basic object on which
the statistical mechanical approach of Lynden-Bell (1967) is based are phase
space elements of {\it constant} volume but of {\it different} mass, it is
found that the mass independence of the equations of motion of Violent 
Relaxation is not fully represented. Considering phase space elements of  {\it
different} volume but of {\it constant} mass removes the velocity  dispersion
problem tendentially present in the investigation of Lynden-Bell (1967).

\section{VIOLENT RELAXATION OF PHASE SPACE ELEMENTS OF DIFFERENT DENSITIES}
The collisionless nature of the process of Violent Relaxation allows
to consider its dynamics in terms of the one-particle phase 
space $\mu$. The Vlasov equation 
\begin{equation}
D_t f\ = \partial_{\bf x} f \cdot {\bf v} -
\vec{\nabla}_{\bf x} \Phi \cdot \partial_{\bf v} f = 0
\label{vlasov}
\end{equation}
governs the collisionless dynamics of the fine-grained phase space density 
$f$ during Violent Relaxation. Violent Relaxation is essentially a 
self-gravitating process. Accordingly, the gravitational potential $\Phi$ 
in (\ref{vlasov}) is given by
\begin{equation}
\Phi({\bf x})=-G\int{\int f({\bf v'},{\bf x'})\,d^3v'
\over{|{\bf x}-{\bf x'}|}}\,d^3x' \, .
\label{poisson}
\end{equation}
Equations (\ref{vlasov}) and (\ref{poisson}) describe the 
dynamics of Violent Relaxation in terms of the continuum limit without 
reference to the individual mass of the objects involved. We will 
restrict ourself to the continuum limit description and assume its 
validity hereafter. As a consequence, the discussion is limited to 
the phase space element approach (Lynden-Bell 1967) and does not 
refer to its counterpart, the particle approach of Shu (1978) 
(see also Madsen (1987) and Shu (1987)).   
 
By considering Violent Relaxation in the terminology of phase space elements,
the attempt of Lynden-Bell (1967) incorporates the constraint on the time 
evolution of the $\mu$-space provided by the Vlasov equation (\ref{vlasov}) 
stating the constancy of the phase space density $f$ along trajectories in 
$\mu$-space. As a consequence of the Vlasov equation, initially not
overlapping  phase elements do never overlap. Therefore, a microscopic
exclusion principle for  phase space elements in $\mu$-space is established.
The second  characteristic feature of Violent Relaxation reflected by equation 
(\ref{vlasov}) is the independence of the equations of motion on the
individual  phase space element or particle masses. It is incorporated into the
statistical  mechanical picture by the dimensionality of the $\mu$-space which
is chosen  to yield the mass density in units mass per spatial and velocity
volumes,  $\Delta^3x$ and $\Delta^3v$, respectively. 

In order to apply statistical mechanical methods, the $\mu$-space is 
divided into a large number of microcells. With $\eta$ the phase space 
density of a phase space element (i.e. an occupied microcell) and 
$\omega$ its volume, the mass associated with the phase element is 
$\eta\omega$. A microstate may then be described by the set of 
occupied microcells $\omega_j$ and their density $\eta_j$. While 
the volume occupied by the microstate in the $\mu$-space is 
\begin{equation}
\Delta \mu = \sum_j N_j \omega_j \, ,
\end{equation}
the corresponding volume $\Delta\Gamma$ in $\Gamma$-space is  
\begin{equation}
\Delta\Gamma=\prod_j \omega_j^{N_j} (\eta_j \omega_j)^{3N_j} 
\label{gammavol1}
\end{equation}
where $N_j$ is the total number of phase space elements of density 
$\eta_j$ occupying microcells of volume $\omega_j$. 
The first crucial point is that the phase space volume in 
$\Gamma$-space (\ref{gammavol1}) not only depend on the volumes
$\omega_j$ but does also depends on the phase space density $\eta_j$. 
The second point concerns the total energy of the microstate
\begin{equation}
E=\sum_{j} \eta_j\omega_j ({1 \over{2}} {\bf v}_j^2 + \Phi_j)
\end{equation}
where ${\bf v}_j$ is the velocity associated with the $i$th microcell, and
$\Phi_j$ is the gravitational potential which the microcell is exposed to. 
What is of importance here is that the total energy $E$ is related to the 
$\Gamma$-space, i.e. it is not possible to derive $E$ independently of 
the absolute value of the phase space density $\eta_j$, or strictly speaking, 
of the mass $\eta_j \omega_j$ of the phase space element. This dependence
is recovered in the macroscopic constraints the system is subject to
in the statistical mechanical approach. In this sense, the basic 
objects on which the Lynden-Bell (1967) statistic is based are not 
the (mass independent) phase space element volumes $\omega_j$ in 
$\mu$-space but their mass dependent counterparts 
$\omega_j (\eta_j \omega_j)^{3}$ in $\Gamma$-space. Therefore the 
discussion does, at least with respect to the energy constraint, 
not reflect the universal mass independence of the equations of 
motion of Violent Relaxation. It is indeed easy to see that it is 
the energy constraint which formally introduces the dependence of 
the coarse grained phase space distribution on the phase space 
density $\eta_j$ in the approach of Lynden-Bell (1967). This means 
that in order to incorporate or represent the mass independence it is 
not enough to just chose the dimensionality of the $\mu$-space as 
discussed above.

We seek for a generalized constraint related to the division of the 
$\mu$-space into phase space elements and microcells, respectively.
This division of the $\mu$-space refers to the intermediate 
step of the determination of the most probable state in terms of occupation 
numbers of macrocells. While this step relies on a specific division of the 
$\mu$-space, the final result is, in contrast, independent of it as will
become clear later. Insofar and with respect to the continuum limit, the 
volumes of the phase space elements are arbitrary and have no direct 
physical significance. Instead, physical significance is attributed to the 
mass $\eta_j \omega_j$ (or the corresponding mass differences) of the
phase space elements in $\mu$-space reflecting the properties of
the basic counterparts $\omega_j (\eta_j \omega_j)^{3}$ in 
$\Gamma$-space. In this respect it is natural to incorporate the 
universal mass independence of the equations of motion of Violent 
Relaxation into the statistical mechanical picture by introducing a 
constraint on the volumes $\omega_j$ stating that the corresponding
phase elements have constant mass, i.e.
\begin{equation}
\eta_j\omega_j = \mbox{const.}
\label{vol1}
\end{equation} 
We note that this procedure is in perfect agreement with the
nature of the microscopic exclusion principle of 
Lynden-Bell (1967) which, for the case of only one
phase space density, states that the mass of the phase elements is
constant. The approach characterized by
(\ref{vol1}) may thus be considered to be just its generalization.
However, as a result and in contrast to Lynden-Bell (1967), where the phase 
space elements have different mass but constant volume, the phase 
space elements considered here differ in  volume while having constant 
mass. 

\section{STATISTICAL MECHANICS OF VIOLENT RELAXATION\label{statinv}}
In this section we attempt to derive the coarse grained phase space 
distribution $\bar{f}$ of the final state of a system consisting
of phase space elements of $J$ different phase space densities $\eta_j$ 
subject to Violent Relaxation. The densities $\eta_j$ are obtained from 
the initial fine grained phase space density $f$. To determine the final 
state, we apply the same maximum entropy principle as used by Lynden-Bell 
(1967). The state which maximizes the entropy under 
the constraint of conserved total energy and mass is considered as 
the most probable final state attained by the system. As has been
pointed out by several authors [e.g. Lynden-Bell (1967), Shu (1978),
Madsen (1987), Shu (1987)], this final state is rather an idealized 
final state since it implicitly relies on the assumption of complete
mixing in phase space which eventually does not occur in real 
systems because of limited time scales.      

Let us start with the determination of the volume $\omega_j$ of a phase 
element of density $\eta_j$. According to (\ref{vol1}), this volume 
is given by
\begin{equation}
\omega_j = {m /{\eta_j}} 
\label{vol2}
\end{equation} 
where $m$ is considered to be constant and is interpreted as the individual 
mass of all phase space elements under consideration. In order to simplify 
the discussion but not introducing further restrictions, we now assume 
that the different phase space volumes $\omega_j$ are all multiples of 
some smallest volume 
\begin{equation}
\omega=\mbox{min}\{\omega_j\} \, .
\end{equation}
The volume $\omega_j$ occupied by a phase space element of density
$\eta_j$ becomes
\begin{equation}
\omega_j= g_j \omega 
\label{vol3}
\end{equation}
where $g_j$ is the factor by which $\omega_j$ is larger than the smallest 
volume $\omega$, i.e. the microcell $\omega_j$ is made up of $g_j$ microcells
of elementary volume $\omega$.

The $\mu$-space is assumed to be divided into a large number of microcells 
of volume $\omega$ each. At the macroscopic level, these microcells
are grouped into macrocells containing a large number $\nu$ of microcells. 
The corresponding volume of the macrocell is $\nu\omega$. Suppose that there 
is a microstate in which the $a$th macrocell contains $n_{aj}$ phase space 
elements of different densities $\eta_j$. Due to the collisionless nature 
of the interaction in the Violent Relaxation process described by the 
Vlasov equation (\ref{vlasov}) there is no cohabitation of microcells 
(Lynden-Bell 1967). 

We now calculate the number of ways of assigning $n_{aj}$ microcells 
of volume $\omega_j$ and density $\eta_j$ without cohabitation to the $a$th 
macrocell. Under the assumption that the phase space is resolved to the
scale of the volume of the smallest microcell $\omega$ we find that the
number of ways of assigning $n_{aj}$ microcells of volume $\omega_j$
without cohabitation to the $a$th macrocell is, according to 
(\ref{vol3}), approximated by ${\nu!/{(\nu - g_j n_{aj} )!}}$.
This number includes cases where the 
microcells of volume $\omega_j$ assigned to the macrocell may not entirely
belong to it. However, these particular situations are rare since their 
relative occurrence is proportional to 
$\approx 1/\sqrt{\nu}$, where $\nu \gg 1$.     
In order to determine the total number of ways of assigning the $\sum_j n_{aj}$ 
phase space elements to the $a$th macrocell we assume the $g_j$ are ordered 
according to their values. Starting with the highest value assigned to $g_1$ or
with the largest phase space element of volume $\omega_1$, respectively, the  
total number of ways of assigning the $\sum_j n_{aj}$ phase space elements to 
the $a$th macrocell is
\begin{eqnarray}
w(n_{aj})&=&{\nu!\over{(\nu - g_1 n_{a1} )!}}\cdot \nonumber \\
& &{(\nu- g_1 n_{a1})!\over{((\nu - g_1 n_{a1}) - g_2 n_{a2})!}}\cdot \nonumber \\
& & \, \ldots \, \cdot \nonumber \\
& & {(\nu- \sum_j^{J-1}g_j n_{aj})!\over{(\nu - \sum_{j}^J g_{j} n_{aj})!}} \nonumber \\ 
&=&{\nu!\over{(\nu-\sum_{j} g_{j} n_{aj} )!}} \, \, .
\label{microdisp}
\end{eqnarray}
By the ordering of the $g_j$ and the exclusion principle expression 
(\ref{microdisp}) does not exclude every situation corresponding to 
partially overlapping phase space elements. Consider for instance a 
situation where a phase space element of volume $\omega_j$ is assigned to
a position between regions of the macrocell which have been already 
occupied. If the volume $\omega_j$ is larger than the volume between 
the occupied regions then a situation is recovered in which phase space 
elements may partially overlap. However, as is the case for phase 
space elements overlapping the macrocell, situations representing overlapping 
phase space elements are also rare. This applies in particular to 
cases characterized by $n_{aj} \omega_j \ll \nu \, , \forall j$. We 
will return to this question in a subsequent section. Here we assume 
that expression (\ref{microdisp}) yields a sufficiently good approximation
to the total number of ways of assigning the $\sum_j n_{aj}$ phase 
space elements to the $a$th macrocell.

The total number of microstates $W(\{n_{aj}\})$ corresponding to a given set 
of occupation numbers $\{n_{aj}\}$ is found by multiplying the expressions
(\ref{microdisp}) and taking into account the number of ways of splitting
the total of $N$ distinguishable elements into groups $n_{aj}$. This 
yields
\begin{eqnarray}
& & W(\{n_{aj}\})
=\prod_j{N!\over{\prod_a\, n_{aj} !}}\cdot \prod_a\,w(n_{aj}) \nonumber \\
& & =\prod_j{N!\over{\prod_a\, n_{aj} !}}\cdot 
\prod_a\,{\nu!\over{(\nu-\sum_{j} g_{j} n_{aj} )!}} \, .
\label{macrodisp}
\end{eqnarray}
Since the numbers $w(n_{aj})$ are approximations, expression (\ref{macrodisp})
has also to be considered as an approximation.

The macroscopic constraints the system is subject to 
are the $J$ constraints related to conservation of 
the total number $N_j$ (or total mass $M_j$) of phase space elements of 
densities $\eta_j$
\begin{equation}
\sum_{a} \eta_j\omega_j n_{aj} = \eta_j\omega_j N_j = M_j\, .
\label{constNJ}
\end{equation} 
The energy constraint reads
\begin{equation}
\sum_{j}\sum_{a} \eta_j\omega_j n_{aj}({1 \over{2}} {\bf v}_a^2 + 
{1 \over{2}} \Phi_a) = E 
\label{constE}
\end{equation}
where the gravitational potential is defined by
\begin{equation}
\Phi_a=\Phi({\bf x}_a)=-\sum_{j} \sum_{b=1, b \not= a}
   {G \eta_j\omega_j n_b \over{\left| {\bf x}_a - {\bf x}_b \right|}} \, .
\end{equation}
Note, that the volumes $\omega_j$ or, respectively, the associated masses
$\omega_j \eta_j$ enter the calculation through these macroscopic 
constraints.  

The most probable state is found by the standard procedure of maximizing
$\log W$ subject to the constraints of constant total energy and constant mass. 
Introducing the Lagrangian multipliers $\alpha_j$ and $\beta$, respectively,
related to the constraints (\ref{constNJ}) and (\ref{constE}) the expression 
to be maximized is
\begin{eqnarray}
\lefteqn{\ln[W(\{n_{aj}\})] - \alpha_j \, \sum_{j}\sum_{a} \eta_j\omega_j n_{aj} - }\nonumber \\
 & & \beta \, \sum_{j}\sum_{a} \eta_j\omega_j n_{aj}({1 \over{2}} {\bf v}_a^2 + 
{1 \over{2}} \Phi_a) \, . 
\label{variation1}
\end{eqnarray}
The variation of (\ref{variation1}) leads to
\begin{eqnarray}
\lefteqn{\sum_{a} \left( -\ln[n_{aj}] + \ln[\nu-\sum_{j} g_{j} n_{aj} ] \right)
\delta n_{aj} -}  \nonumber \\
& & \alpha_j \sum_{a} \eta_j\omega_j \delta n_{aj} - \\ \nonumber
& & \beta \sum_{a}  \eta_j\omega_j \left(\left({1 \over{2}} {\bf v}_a^2 + 
{1 \over{2}}\Phi_a \right)\delta n_{aj} + {1 \over{2}}n_{aj} 
\delta \Phi_a \right) \nonumber \\
&=& 0 \nonumber \, 
\label{variation2}
\end{eqnarray}
and by taking into account the independence of the $\delta n_{aj}$'s, 
equation (\ref{variation2}) splits into the $J$ equations 
\begin{eqnarray}
\ln\left[{\nu-\sum_{j} g_{j} n_{aj}  \over{n_{aj}}}\right]= 
\alpha_j \eta_j\omega_j  +  \beta \eta_j\omega_j \epsilon_a \, , \\
 \quad (j=1, \ldots , J) \nonumber
\label{bed1}
\end{eqnarray}
where $\epsilon_a={1 \over{2}} {\bf v}_a^2 + \Phi_a$ stands for the 
total energy per unit mass related to the $a$th microcell. 
The corresponding occupation numbers $\{n_{aj}\}$ are
\begin{equation}
n_{aj}=(\nu-\sum_{j} g_{j} n_{aj} )\exp[-\alpha_j \eta_j\omega_j  -
\beta \eta_j\omega_j \epsilon_a] \, .
\label{bed2}
\end{equation}
By multiplying (\ref{bed2}) with $g_j$, summing over $j$ and subsequent
algebraic manipulations we find
\begin{eqnarray}
\lefteqn{(\nu -{\sum_{j} g_{j} n_{aj}}) =}  \\
& &{\nu \over {\sum_{j} g_j \exp[-\alpha_j \eta_j\omega_j  -
\beta \eta_j\omega_j \epsilon_a] +1}} \nonumber\, .
\end{eqnarray}
According to (\ref{bed2}), with $\mu_j=-\alpha_j/\beta$ and  
with the constant $\omega_j \eta_j = m$ from equation (\ref{vol2}) 
the most probable occupation numbers $\{n_{aj}\}$ become
\begin{equation}
n_{aj}={\nu\exp[-\beta m (\epsilon_a-\mu_j)]
\over{\sum_{j} g_j \exp[-\beta m (\epsilon_a-\mu_j)] +1}} \, .
\label{lbequ2a}
\end{equation}
The coarse grained phase space distribution $\bar{f}$ is defined as 
the sum of the $J$ phase space distributions $\bar{f_j}$ where
\begin{equation}
\bar{f_j}({\bf v},{\bf x}) \approx \bar{f_j}({\bf v}_a,{\bf x}_a) 
={g_j n_{aj} \eta_j \over{\nu}} \, .
\label{fbarscale}
\end{equation}
It is thus given as 
\begin{eqnarray}
\bar{f}({\bf v},{\bf x}) &\approx& \bar{f}({\bf v}_a,{\bf x}_a) \nonumber \\
&=&\sum_{j} \bar{f_j}({\bf v}_a,{\bf x}_a) \nonumber \\
&=&\sum_{j} {g_j n_{aj} \eta_j \over{\nu}} \, .
\end{eqnarray}
Substituting for $\eta_{aj}$ from (\ref{lbequ2a}), the coarse grained 
phase space distribution $\bar{f}$ becomes finally
\begin{eqnarray}
\lefteqn{\bar{f}({\bf v},{\bf x})=}  \\
& &\sum_{j} {g_j \eta_j  \exp[-\beta m (\epsilon({\bf v},{\bf x})-\mu_j)] 
\over{\sum_{j} g_j \exp[-\beta m (\epsilon({\bf v},{\bf x})-\mu_j)] +1}}\nonumber
\label{lbequ2}
\end{eqnarray}
where $\epsilon({\bf v},{\bf x}) = {1 \over{2}} {\bf v}^2 + \Phi({\bf x})$
is the total energy per unit mass.
According to the initial assumptions, the state characterized 
by the occupation numbers (\ref{lbequ2a}) or the coarse grained 
phase space distribution (\ref{lbequ2}), respectively, is identified 
with the (idealized) final state attained by a system consisting of
phase space elements of different densities $\eta_j$ and which is 
subject to Violent Relaxation. The Fermi-like functional form of 
the respective distributions (\ref{lbequ2a}) and (\ref{lbequ2}) 
originates from the exclusion principle which has been applied 
in order to exclude situations corresponding to cohabitated microcells.   

One should keep in mind, that the distributions (\ref{lbequ2a}) and 
(\ref{lbequ2}), respectively, have been derived under the 
approximations made in evaluating the total number of microstates
$W(\{n_{aj}\})$ corresponding to a given set of occupation numbers 
$\{n_{aj}\}$. However, for the two limiting cases 
($\bar{f_j} \ll \eta_j \, , \forall j$) and
($\bar{f_j} \approx \eta_j \, , \exists j$) the approximation
made in (\ref{microdisp}) is very good since for both situations
the overlap of phase space elements of different volumes
$\omega_j$ does not occur. 

In finishing this section let us return to the question of the explicit
determination of the microcell and macrocell volumes $\omega_j$ and
$\nu \omega$, respectively. While the volumes $\omega_j$ are absorbed in
the common constant mass factor $m=\omega_j \eta_j$ due to relation 
(\ref{vol2}), equations (\ref{lbequ2a}-\ref{lbequ2}) illustrate by the 
canceling of $\nu$ the arbitrariness of the volume $\nu \omega$ of 
the macrocell.
The only constraint thus remaining is related to the applicability of 
statistical methods, i.e. the constraint $\nu \gg 1$. 

\section{DISCUSSION AND CONCLUSIONS}
In order to discuss the result (\ref{lbequ2}) we compare the equilibrium 
distribution $\bar{f}$ with the coarse grained phase space distribution 
$\bar{f}_{LB}$ which was originally derived by Lynden-Bell (1967)
\begin{eqnarray}
\lefteqn{\bar{f}_{LB}({\bf v},{\bf x})=} \\
& &\sum_{j} {\eta_j  \exp[-\beta_j (\epsilon({\bf v},{\bf x})-\mu_j)] 
\over{\sum_{j} \exp[-\beta_j (\epsilon({\bf v},{\bf x})-\mu_j)] +1}} 
\nonumber \, .
\label{lbequ3}
\end{eqnarray}
In the non-degenerate limit characterized by 
$\bar{f_j} \ll \eta_j \, , \forall j$, the
coarse grained phase space distribution (\ref{lbequ3})
turns into a sum of Maxwellians of different velocity dispersions 
$\sigma_j^2=\beta_j^{-1}$
\begin{equation}
\bar{f}_{LB}({\bf v},{\bf x})=\sum_{j} 
\eta_j  \exp[-\beta_j (\epsilon({\bf v},{\bf x})-\mu_j)] 
\label{lboltz1}
\end{equation}
where $\beta_j \propto \eta_j$.
In the non-degenerate limit, the final state of the process of Violent 
Relaxation is therefore identified with a thermalized state. This fact 
has been criticized by Madsen (1987) and Shu (1987). The velocity 
dispersions of the individual components are inversely proportional 
to their phase space densities and, as a consequence, the components 
have the same temperature, a fact one would expect for a system subject 
to a two-body relaxation process. This is not only counterintuitive, 
but it is also in disagreement with observations of astrophysical objects
which are believed to have been subject to Violent Relaxation.

In contrast, the corresponding result derived form (\ref{lbequ2}) 
turns out to reflect the independence of the Violent Relaxation process 
on the phase space densities $\eta_j$ of the different components involved.
In the non-degenerate limit ($\bar{f_j} \ll \eta_j \, , \forall j$), 
the coarse grained phase space distribution (\ref{lbequ2}) does again 
become a sum of Maxwellians. But the Maxwellians are all characterized 
by {\it one} single velocity dispersion $\sigma^2=\beta^{-1}$. The coarse
grained phase space distribution in the non-degenerate limit is 
\begin{equation}
\bar{f}({\bf v},{\bf x})=\sum_{j}
 g_j \eta_j  \exp[-\beta m (\epsilon({\bf v},{\bf x})-\mu_j)]  \, .
\label{lboltz2}
\end{equation}
We note, that according to Section \ref{statinv} the approximation
of $W(\{n_{aj}\})$ on which (\ref{lboltz2}) is based is very good, because
the condition $n_{aj} \omega_j \ll \nu \, , \forall j$ is equivalent to
the condition $\bar{f_j} \ll \eta_j \, , \forall j$. In the non-degenerate 
limit, the final state of the Violent Relaxation process defined by 
(\ref{lboltz2}) is the superposition of Maxwellians of a common 
velocity dispersion. This is in agreement with the common velocity 
dispersion of the different components observed for instance
in clusters of galaxies (e.g. Lubin \& Bahcall 1993). The common 
velocity dispersion is equivalent to an equipartition of energy per 
mass. As a consequence, there is no mass-segregation present in systems 
subject to Violent Relaxation as has also been observed. 

Another limiting case for which the approximation made in evaluating the coarse
grained phase space distribution (\ref{lbequ2}) yields an exact result is found
for  ($\bar{f_j} \approx \eta_j \, , \exists j$). This situation corresponds to
a dense coverage of the phase space by phase space elements with one specific
density $\eta_j$.  

For the case of dense coverage of the phase space by phase 
space elements of different volumes $\omega_j$ 
(i.e. $\bar{f_j} \,^<\!\!\!\!_\sim\, \eta_j \, , \forall j$) 
the situation becomes more complicated because the approximations 
which led to equation (\ref{macrodisp}) worsen. It is, however, clear that 
the effective total number of microstates corresponding to a given 
set of occupation numbers $\{n_{aj}\}$ is less than $W(\{n_{aj}\})$ 
since, as discussed in Section \ref{statinv}, the number $W(\{n_{aj}\})$ 
includes cases which correspond to partially overlapping phase space 
elements. The question remains open of how the correction of this 
approximation would influence our result. An investigation of this 
point is left for subsequent work.

In summary, we have performed a statistical mechanical investigation
of Violent Relaxation of phase space elements of different mass densities. 
A well motivated constraint on the volume of the phase space elements has 
been introduced stating all phase space elements independent of their 
density to be of constant mass. As a consequence, the statistical 
mechanics accounts for the independence of the equations of motion
in the process of Violent Relaxation on the mass density of the phase 
elements. As in the case of Lynden-Bell (1967), the final state reached 
by a system which is subject to Violent Relaxation is characterized by 
a Fermi-like coarse grained phase space distribution $\bar{f}$ reflecting 
the microscopic exclusion principle (Lynden-Bell 1967) that has been 
applied in order to account for the collisionless nature of Violent
Relaxation. However, the functional form of this coarse grained phase 
space distribution is independent of the different phase space densities
of the interacting components. In the non-degenerate limit, the  
coarse grained phase space distribution $\bar{f}$ becomes a superposition
of Maxwellians, $\bar{f}_i$, all possessing the same velocity dispersion. 
This is in agreement with the observation of a common velocity dispersion 
of different components of astrophysical objects which are believed to
have undergone Violent Relaxation (e.g. the different matter components 
in clusters of galaxies). Thus, the velocity dispersion problem which 
was present in the original treatment of Violent Relaxation 
(Lynden-Bell 1967) has been removed. 

As has been pointed out for a long time (e.g Lynden-Bell 1967, Shu 1978, Madsen
1987) and confirmed by N-body simulations (e.g Melott 1983, McGlynn 1984, White
1996), Violent Relaxation eventually fades before  the final state (eq.
[\ref{lbequ2}]) is attained throughout a real  astrophysical systems. However,
expression (\ref{lbequ2}) is likely to hold in the central regions where
Violent Relaxation occurs most violently. In comparing the results obtained
here with results from N-body simulations,  one should probably take into
account recently expressed scepticism concerning the reliability of N-body
simulation. As has been shown (e.g. Kuhlman et al. 1996, Madsen 1996), there
possibly exists a problem for high-resolution N-body simulations due to
collision and discreteness errors. When dealing  with Violent Relaxation of
nonbaryonic dark matter as, e.g., massive neutrinos (Kull et al. 1996), this
problem becomes worse since  massive neutrinos behave like an almost continuous
medium.         

The assumptions and approximations on which the present investigation 
relies on are (1) that the process of Violent Relaxation can be described 
in the continuum limit, (2) that Violent Relaxation of a multi-component system 
is essentially a collisionless process leading to a final equilibrium state,
(3) that the characteristics of the Vlasov equation in a statistical mechanical
picture of Violent Relaxation have not only to be represented by a 
microscopic exclusion principle but also by considering phase space elements 
of constant mass, and (4) that the approximations made in deriving the 
occupation numbers are valid at least in the non-degenerate case. As a
consequence of condition (1), the present investigation in terms of phase 
space elements does not refer to its counterpart formulated in terms of 
particles (Shu 1978). The main difference compared to Lynden-Bell (1967) 
is the generalized constraint on the mass of the phase elements. 

The assumptions and approximations made here do not 
allow to consider the important case of incomplete Violent
Relaxation. They also do not apply to the fact, that the final state 
reached by Violent Relaxation is, in a global picture, a transient state 
since two-body collisions will eventually start driving the system towards 
a thermalized state. In addition, the present investigation does eventually 
not apply, as discussed, to all cases of dense coverages of the phase space by 
phase space elements. However, we believe that a more sophisticated treatment 
of the problem which corrects for these approximations will not change 
our basic result.


\begin{references} 
\reference{KANDERUP87} Kandrup, H. E. 1987,  \mnras, 225, 995
\reference{KULL96} Kull, A., Treumann, R. A., \&  B\"ohringer, H. 1996,  \apjl,
466, L1 
\reference{LUBIN94} Lubin, L. M., \& Bahcall, N. A.
1993,  \apj, 415, L17
\reference{LBELL67} Lynden-Bell, D. 1967,  \mnras, 136, 101
\reference{MADSEN87} Madsen, J. 1987,  \apj, 316, 497
\reference{MCGLYNN84} McGlynn, T. A. 1984, \apj, 281, 13
\reference{MELOTT83} Melott, A. L. 1983, \apj, 264, 59
\reference{SHU78} Shu F. H. 1978,  \apj, 225, 83
\reference{SHU87} Shu F. H. 1987,  \apj, 316, 502
\reference{SOKER96} Soker, N. 1996,  \apj, 457, 287
\reference{STIAVELLI87} Stiavelli, M., \& Bertin, G. 1987,  \mnras, 229, 61 
\reference{TREMAINE86} Tremaine, S., Henon, M., \& Lynden-Bell, D. 1986,
\mnras, 219, 285
\reference{WHITE96a} White, S. D. M. 1996, to appear in Proceedings of the 
36th Herstmonceux Conference, eds. Lahav, O., Terlevich, E., 
\& Terlevich, R., ({\tt astro-ph/9602021}) 
\end{references}
\end{document}